# Holographic Mapping of Orbital Angular Momentum Using a Terahertz Diffractive Optical Neural Network

Wei Jia, Miguel Gomez, Steve Blair, and Berardi Sensale-Rodriguez
Department of Electrical and Computer Engineering, The University of Utah, Salt Lake City, UT, 84112, USA

**Abstract**: Using orbital angular momentum (OAM) in the terahertz (THz) range provides a new degree of freedom for communication and imaging systems. This study presents a compact diffractive optical neural network designed to recognize discrete and superposed OAM states at THz frequencies. The network consists of six diffractive layers trained to spatially separate nine OAM modes with topological charges from 1 to 9. Each mode is projected to a distinct position on the output plane, enabling direct recognition of its state. The structure was fabricated through low-cost 3D printing techniques with high-impact polystyrene (HIPS), allowing for scalable and practical implementations. Experimental validation at 0.3 THz demonstrates good fidelity of mode discrimination and mapping. The proposed approach offers a robust and economical pathway for OAM decoding, offering new opportunities for beam manipulation through THz systems based on diffractive optical neural networks.

1. Introduction

The terahertz (THz) frequency region has become increasingly relevant for communications, imaging, and sensing. One of its attractive features is the possibility of generating electromagnetic waves carrying orbital angular momentum (OAM), which adds a spatial degree of freedom to the wavefront [1]. The incorporation of OAM into THz systems offers the potential for multi-dimensional data encoding and improved channel capacity via the orthogonality of OAM modes through spatial-mode multiplexing, enabling multi-mode, polarization-division, and multi-user multiplexing in THz links [2–6]. Despite growing interest in OAM-related THz applications, practical implementation remains difficult [1,6]. Some existing methods for detecting OAM states in the THz range rely on bulky setups and complex components that are not easily scalable. Additionally, material losses and fabrication constraints at the THz range further limit the development of compact and efficient OAM mode detection systems [7].

To this end, we adopted a novel THz diffractive optical neural network (DONN) [8], designed to recognize and spatially resolve individual OAM states. The network is capable of detecting nine distinct OAM modes with topological charges (TCs) from 1 to 9, as well as superposed OAM modes. The DONN consists of six inverse-designed diffractive layers that project each OAM mode onto a predefined spatial location at the imaging plane, enabling efficient and direct mode discrimination without the need for complex interferometric techniques.

In the work, high-impact polystyrene (HIPS) is used as the fabrication material because of its low absorption at the desired THz frequencies and excellent compatibility with filament-based 3D printing techniques [9]. In contrast, many commonly used 3D printing materials exhibit high THz absorption losses [9, 10], although materials such as paraffin wax and HDPE show relatively low absorption. In our previous work [11], we demonstrated a low-loss paraffin wax DONN for THz OAM digital display. In general, diffractive optical elements (DOEs) emerged as a versatile platform for wavefront shaping [12–17]. These consist of planar components engineered to manipulate the phase of electromagnetic waves. The use of 3D printed HIPS supports scalable and repeatable fabrication of complex THz DOEs, making it suitable for practical deployment and future customizations [18].

In this work, the design methodology, fabrication process, and experimental validation of the proposed THz DONN are detailed. Utilizing a 0.3 THz transmitter and a THz imaging camera, the reliable detection of individual OAM modes, as well as superposed OAM modes, is demonstrated. The results underscore the potential of this architecture for enabling cost-effective and compact solutions in THz communications and imaging systems.

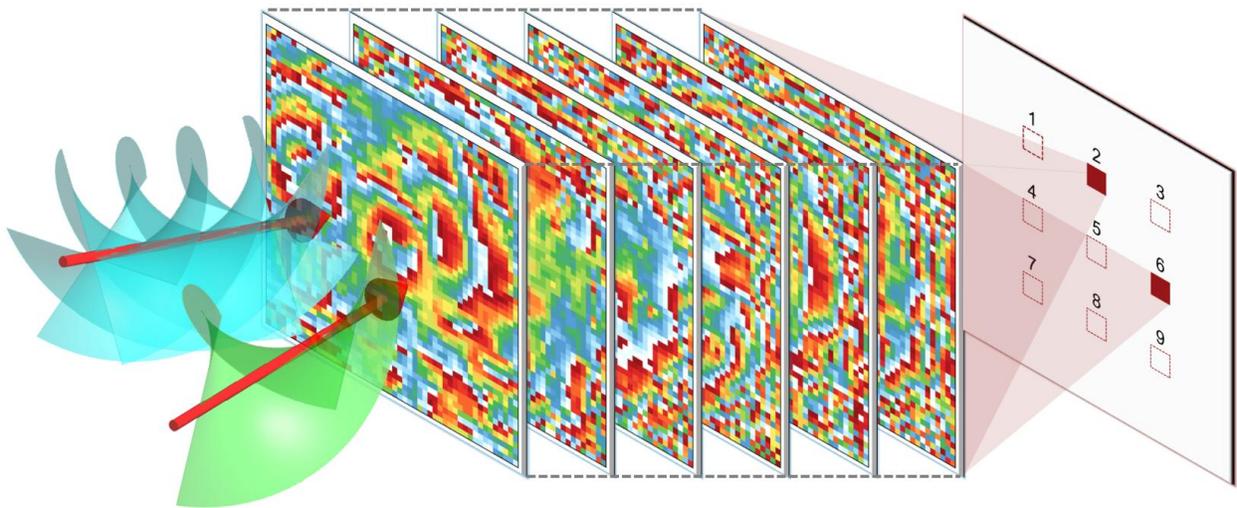

**Fig. 1.** Schematic of the proposed 6-layer diffractive optical neural network for spatial mapping of OAM beam with individual and superposed states.

## 2. Design and Simulation

The schematic of the DONN is shown in Fig. 2(a). The OAM beam is generated by a spiral phase plate (SPP) illuminated by radiation from a 0.3 THz transmitter (VDI). Initially, the beam propagates a distance $d_1$ before reaching the first diffractive layer (DL) of the DONN. The separation between adjacent DLs is denoted as $d_2$. After passing through the final layer, the output beam propagates an additional distance of $d_3$ to reach the imaging plane. Within this system, the SPP serves as the input layer, encoding the OAM beams for processing by the DONN. The sequence of multiple DLs functions as the hidden layers, and the imaging plane serves as the output layer, where the diffracted optical field intensity is characterized. Each point on the diffractive layer functions like a neuron by applying a learned phase shift to modulate the OAM beams from the SPP [8]. It introduces constructive and destructive interference across the diffractive layers to redirect the optical energy to specific spatial locations at the imaging plane.

In the design, each DL has a uniform size of 50 mm x 50 mm, which is close to the dimensions of the utilized THz camera. The layers are discretized into 50 x 50 pixel grids, resulting in a pixel size of 1 mm x 1 mm. Each pixel is a diffractive element, and its phase value $\phi$ is a trainable parameter, constrained between 0 to $h_m$, and is optimized during the training process, where $h_m$ is the maximum pixel height corresponding to 2π phase shift of the material at 0.3 THz. The spacing distance $d_1$, $d_2$ and $d_3$ are 20 mm (~20λ), 20 mm (~20λ), and 50 mm (~50λ), respectively. To more accurately model wave propagation and accommodate a broader frequency bandwidth and sampling, each pixel is subdivided into a 4 x 4 grid of sub-pixels while retaining the same phase value. A zero-padding of 100 pixels is added around the matrix, resulting in a total matrix size of 400 x 400 for each DL. The transmission coefficient of each pixel on the k-th DL is: $t_k(x, y) = exp\,(j\phi_k)$, where $\phi_k = 2\pi/\lambda \cdot h_k \cdot (n-1)$, $\lambda \approx 1$ mm is the wavelength of the 0.3 THz radiation wave, $h_k$ is the actual pixel height corresponding to the $\phi_k$, and $n$ is the refractive index of the utilized material ($n$ = 1.54 at 0.3 THz).

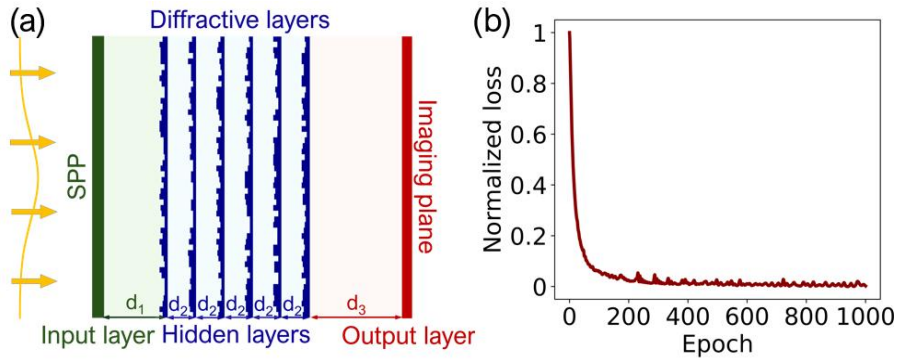

**Fig. 2.** (a) 2D schematic of the diffractive optical network. (b) Normalized training loss as a function of epoch.

To achieve the full 2π phase shift, the maximum pixel height of 1.85 mm is required.

The diffraction field is modeled with the Rayleigh-Sommerfeld diffraction formula, and its propagation kernel is expressed as follows [19]:

$$w(x, y, d) = \frac{d}{r^2}\left(\frac{1}{2\pi r} + \frac{1}{j\lambda}\right) exp\left(\frac{j2\pi r}{\lambda}\right) \qquad (1)$$

where $r = \sqrt{(x'-x)^2 + (y'-y)^2 + d^2}$ and $d$ is the propagation distance. The diffraction field distribution right before the $k$-th layer is $U_k(x, y, z_k)$, and after the modulation of the diffractive layer $t_k(x, y)$, the transmitted complex diffraction field distribution is $T(x, y, z_k) = U_k(x, y, z_k) \cdot t_k(x, y)$. With propagation distance $d$, the field distribution $U_{k+1}$ right before the (k+1)-th layer is obtained by:

$$U_{k+1}(x', y', z_{k+1}) = w(x, y, d) * T(x, y, z_k) = \mathcal{F}^{-1}\{\mathcal{F}\{w(x, y, d)\} \cdot \mathcal{F}\{T(x, y, z_k)\}\} \qquad (2)$$

To accelerate computation, fast Fourier transform (FFT) is utilized, and $F$ and $F^{-1}$ are the FFT and its inverse. The loss function $L_{loss}$ is defined as follows:

$$L_{loss} = \frac{1}{M}\sum_{1}^{M} |I_{opt(m)} - \alpha I_{tgt(m)}|^2 \qquad (3)$$

where $I_{opt(m)}$ represents the diffracted intensity distribution at the imaging plane during each optimization epoch, $I_{tgt(m)}$ denotes the corresponding target intensity distribution at the imaging plane, $M$ is the total number of SPPs, and $\alpha$ is the weighting coefficient, defined as the ratio of the total incident source intensity to the total target intensity of each SPP. The total 9 SPP phase distributions and their corresponding targets at the imaging plane are shown in Fig. 3. Each square in the target is 3 mm × 3 mm in size, corresponding to a 2×2 pixel area on the THz imaging camera.

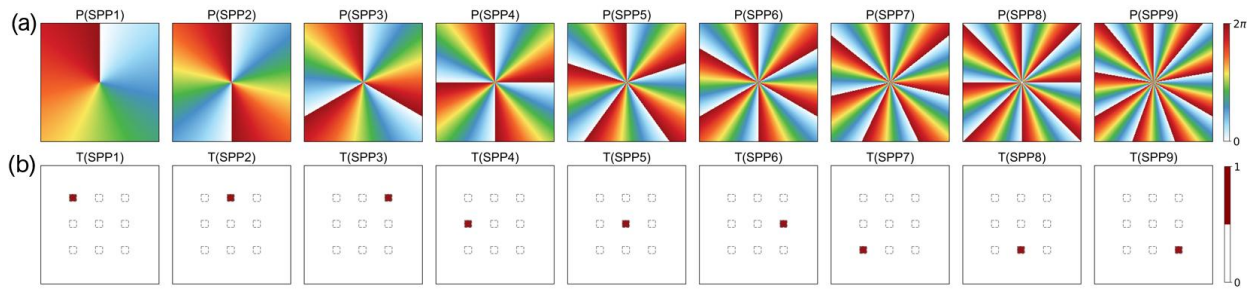

**Fig. 3.** (a) SPP phase distributions with TCs from 1 to 9. (b) Corresponding targets at the imaging plane.

Python and TensorFlow are utilized to implement the training of the DONN. The Adam optimizer [20] drives the learning process by adjusting the phase distribution of the diffractive layers through error backpropagation, with the goal of minimizing the above-defined loss function over successive iterations. The training process is set to run for 1000 epochs with a learning rate of 0.2 without exponential decay. As illustrated in Fig. 2(b), the loss decreases rapidly during the initial stage and gradually saturates in later epochs. Training takes less than a minute using TensorFlow with GPU acceleration on a 3.2 GHz 12th Gen Intel Core i9-12900K and an NVIDIA RTX A2000. The final trained height distribution of each DL of the DONN is shown in Fig. 4(a).

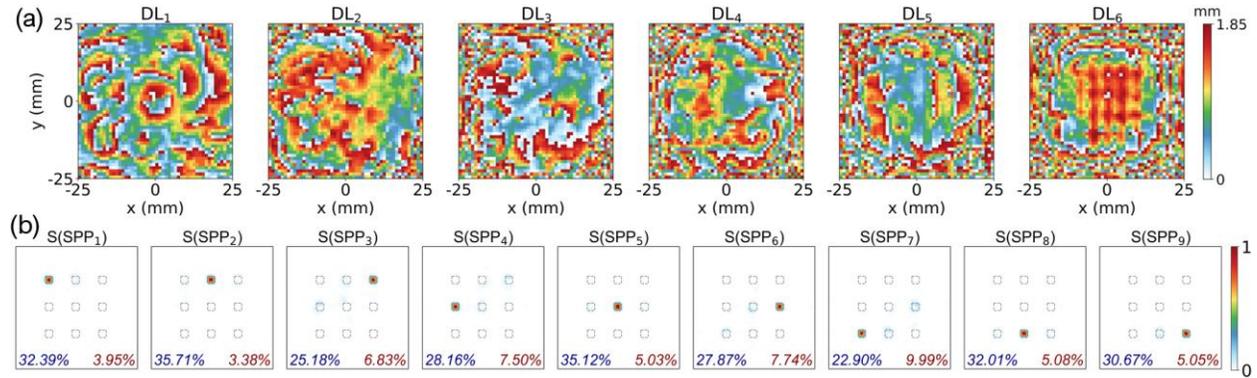

**Fig. 4.** (a) Designed height distribution of each diffractive layer. (b) Normalized simulated intensity distribution at the imaging plane for SPPs with TCs from 1 to 9, bottom left conversion efficiency, bottom right crosstalk.

The simulated intensity distributions at the imaging plane for the nine SPPs are shown in Fig. 4(b). For each SPP, it is evident that the 6-layer DONN efficiently maps the corresponding topological charges of the SPPs to their desired positions. The average conversion efficiency across these 9 SPPs is 30%, demonstrating the DONN high throughput and mapping capabilities. Here, conversion efficiency is defined as the ratio of the total power collected at the desired target spot to the total incident power applied to the SPP. Target-based crosstalk is defined as the ratio of the sum of power leaking into undesired spots to the power at the desired target spot. The simulated crosstalk for each SPP is annotated at the bottom-right of each distribution in red. It is observed that the crosstalk values are consistently low, with an average crosstalk of 6%, indicating good spatial separation of the target spots and high fidelity of mapping performance of the DONN.

### 3. Fabrication and Measurement

To fabricate the designed diffractive network, an Original Prusa MK4S 3D Printer was utilized with HIPS filament. The printer is configured with a 0.25 mm nozzle, while the print settings include a 0.07 mm layer height, 100% infill, and a brim to prevent edge warping. The nozzle and print bed temperatures were set to 245°C and 100°C, respectively. A 1 mm base height is included on each DL to ensure reliable adhesion to the print bed and provide structural support for fine features. Additionally, 3 mm margins were included around each layer to prevent edge deformation and facilitate easier mounting. The 3D printed DLs of the DONN with HIPS filament are shown in Fig. 5(a). To evaluate printing height accuracy, an Olympus

LEXT OLS5000 3D laser confocal scanning microscope was utilized to produce high-precision 3D surface reconstructions, as shown in Fig. 5(b). Pixel 15 is chosen as the base height to align with the designed height. The comparison between measured and designed pixel height is shown in Fig. 5(c). It can be observed that the printed pixel height is very close to the designed pixel height, with a standard deviation of 15 μm, indicating good printing accuracy. To generate OAM beams with TC from 1 to 9, the corresponding SPPs are also printed with the same printing settings, and the photographs of the printed SPPs is shown in Fig. 5(d).

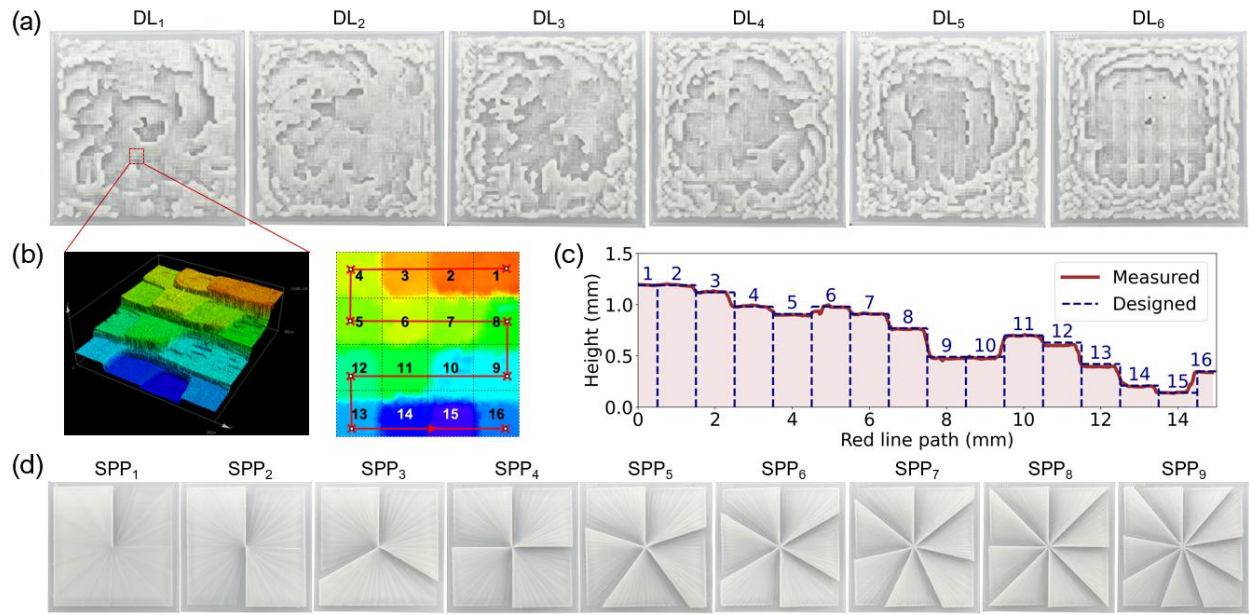

**Fig. 5.** (a) 3D printed six diffractive layers of DONN. (b) Pixel height characterization with a confocal scanning microscope. (c) Comparison between the design and measured height. (d) 3D printed SPPs with TCs ranging from 1 to 9.

The schematic and photograph of the experimental characterization setup are shown in Fig. 6(a). A 0.3 THz beam is emitted from the transmitter and collimated using two printed Fresnel lenses, $FL_1$ and $FL_2$. The normalized simulated intensity distributions in the x-z plane for both lenses are presented in Fig. 6(b). $FL_1$ has a diameter of 80 mm and a focal length of 160 mm, while $FL_2$ has a diameter of 100 mm and a focal length of 200 mm. A photograph of the 3D-printed Fresnel lenses with HIPS material is shown in Fig. 6(c), and a 5 mm margin size is added for easier mounting. The spacing between the two lenses is tuned to achieve a collimated beam. The incident beam to the SPP was the collimated 0.3 THz beam, which passes through the SPP to generate corresponding OAM beams. These beams then served as the input to the 6-layer diffractive network. A Tera-1024 THz camera was utilized to capture the resulting intensity patterns at the imaging plane, which is 50 mm away from the last DL of the DONN. The camera features a 32 x 32 pixel array with a 1.5 mm pixel pitch. An optical rail was utilized to hold the components to ensure precise alignment. 3-axis translational stages are utilized to further fine-tune the positions of the SPP and DONN to achieve better alignment.

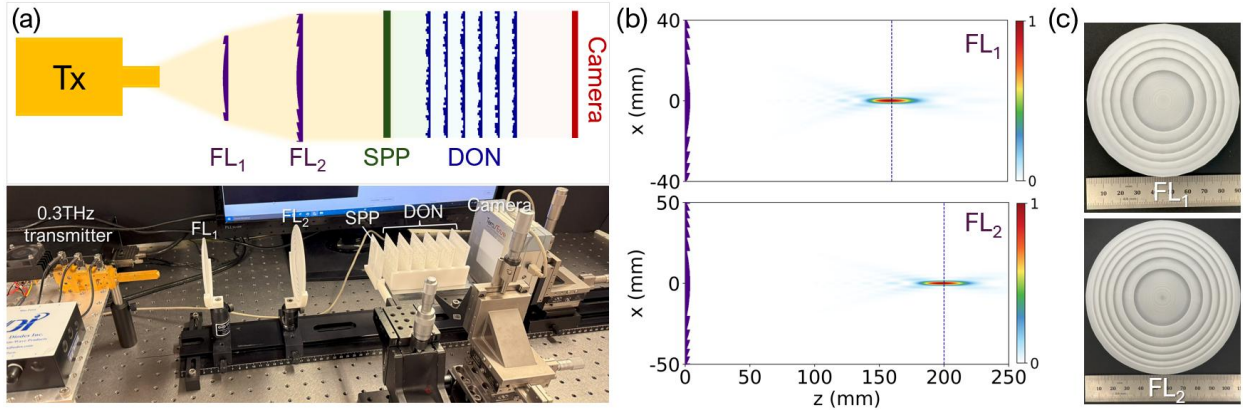

**Fig. 6.** (a) Schematic and photography of experimental characterization setup. (b) Simulated intensity distribution in x-z plane of two designed Fresnel lenses. (d) 3D printed Fresnel lenses with HIPS material.

### 4. Results and Discussion

The measured intensity distributions at the imaging plane for the 9 SPPs are shown in Fig. 7. These intensity distributions were captured with the previously mentioned Tera-1024 THz camera at a distance of 50 mm away from the last layer of the DONN. During the measurement, the camera is configured with the following settings: White point 61, Black point 38, Gamma 50, and Exposure 4. As shown in Fig. 7, it is evident that for each SPP, the DONN can successfully map the corresponding topological charge to the desired spatial locations at the imaging plane. A related measurement video with different SPPs is shown in the Supplementary Information.

Although the intensity distributions confirm the intended spatial mapping of TCs overall, small deviations are observed. These arise primarily from unavoidable alignment tolerances between the SPP, DONN, and camera along the optical axis, leading to slight distortions in the recorded intensity profiles. This affects measurement fidelity, as shown in [21]. Mechanical drift in the setup can also introduce angular offsets of the beam. In addition, phase errors from the 3D printing process, limits of the SPP and DONN, and response uniformity of the Tera-1024 camera further contribute to the measured deviations.

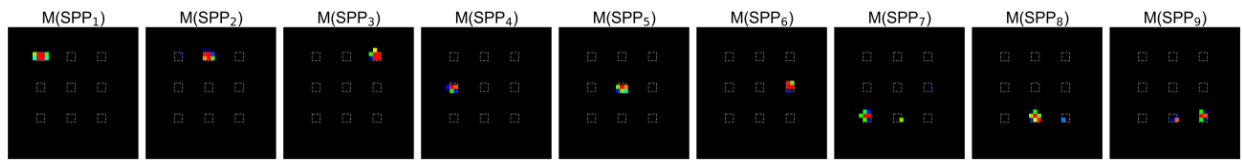

**Fig. 7.** Measured intensity distributions at the imaging plane for the 9 SPPs with TCs ranging from 1 ~ 9.

To analyze the robustness of the designed DONN, intensity distributions are simulated with variations in the imaging plane distance within the range of $d_3 \pm \Delta = 50$ mm $\pm 3$ mm, as illustrated in Fig. 8(a). $SPP_5$ is chosen to evaluate the performance of the diffractive network to the changes in the imaging plane distance. Their corresponding conversion efficiency and crosstalk are shown in Fig. 8(b) over a distance range from 47 mm to 53 mm with a step size of 1 mm. The results indicate that the maximum conversion efficiency occurs at the target imaging distance of 50 mm. As the distance deviates from this optimum, the conversion efficiency decreases progressively: the larger the deviation, the greater the reduction. A similar trend is observed in crosstalk behavior. At the target distance, the DONN directs the THz beam effectively, resulting in enhanced total power throughput. While this maximizes energy transfer to the intended spatial locations, it also increases the likelihood of energy coupling into neighboring locations, thereby elevating crosstalk. As the imaging distance shifts away from the optimal position, the network's ability to direct energy to that distance degrades, leading to a concurrent reduction in both conversion efficiency and crosstalk.

The intensity distributions at different imaging distances are shown in Fig. 8(d). At the target distance of 50 mm, the intensity profile is the most uniform and accurately reproduces the target square pattern, with a deviation of only ±1 mm. When the distance deviates by more than 2 mm, the intensity remains concentrated in the intended region but exhibits shape distortion and non-uniform distribution, and the square geometry is no longer preserved. Nonetheless, the energy still focuses on the intended spatial locations, though with greater distortion. The Pearson correlation coefficient (PCC) was calculated at each distance to quantify the similarity between the simulated and target intensity distributions. As shown in Fig. 8(c), the PCC reaches 0.90 at 50 mm, indicating a close match to the target profile. With a ±1 mm deviation, the PCC slightly decreases to 0.89, still representing a good match. For larger deviations, the PCC declines more significantly, indicating a poorer match—consistent with the visual observations in Fig. 8(d).

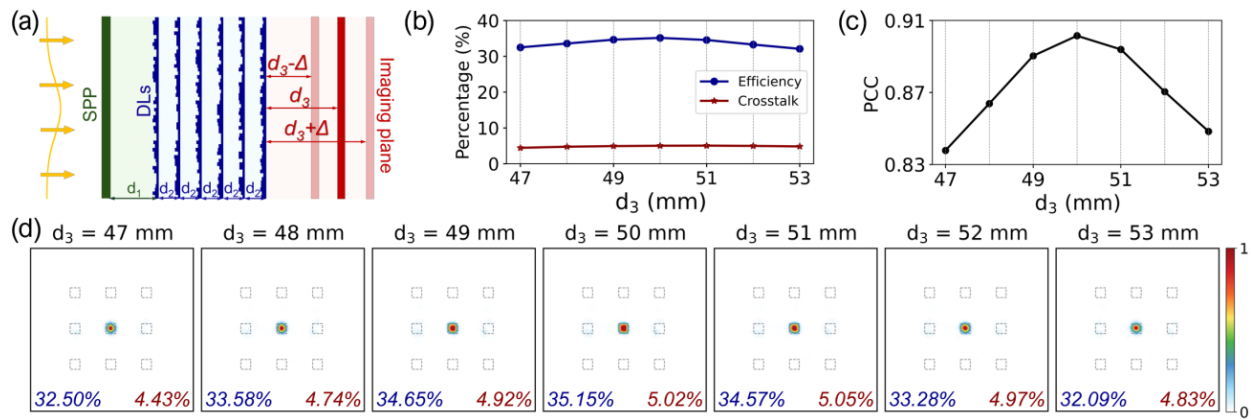

**Fig. 8.** (a) Schematic of sensitivity analysis with variation in imaging distance $d_3 \pm \Delta$. (b) Conversion efficiency and crosstalk of $SPP_5$ at different imaging distances. (c) PCC values versus imaging distance. (d) Imaging plane intensity distributions of $SPP_5$ under varied $d_3$.

The influence of the 3D printing accuracy on the performance of the diffractive network is also investigated. In cases where the height of each pixel on the diffractive layers is uniformly shifted, such as due to variations in the 3D printer's layer height settings, this introduces a global phase offset across the entire network. Since the output intensity distribution is determined by the relative phase difference rather than the absolute phase, such uniform shifts have minimal impact on the performance of the diffractive network.

Random fabrication errors were modeled as Gaussian-distributed pixel height perturbations, reflecting the aggregate effect of possible small, independent sources of error such as mechanical jitter, thermal drift, and platform instability. By the central limit theorem, these effects are well-approximated by a normal distribution. The schematic of the sensitivity analysis for pixel height variations is shown in Fig. 9(a). In the model, zero-mean Gaussian perturbations $\epsilon(x,y) \sim N(0, \sigma^2)$ were added to the designed pixel heights $h(x,y)$ of each diffractive layer, producing perturbed heights $\tilde{h}(x,y) = h(x,y) + \epsilon(x,y)$.

The variation of conversion efficiency and crosstalk of $SPP_4$ as a function of noise standard deviation $\sigma$ (0 ~ 180 µm) is shown in Fig. 9(b). For $\sigma < 30$ µm, the conversion efficiency remains nearly unchanged. As $\sigma$ increases, the efficiency progressively decreases, accompanied by a decrease in crosstalk. The PCCs remain largely constant for $\sigma < 120$ µm, as shown in Fig. 9(c), indicating that the spatial intensity distribution within the target region is preserved, as confirmed by the corresponding intensity distributions in Fig. 9(d). The efficiency-to-crosstalk ratio (ECR) decreases with increasing $\sigma$, indicating that pixel height noise degrades overall performance. Nevertheless, the diffractive network demonstrates strong robustness to fabrication noise for $\sigma < 30$ µm.

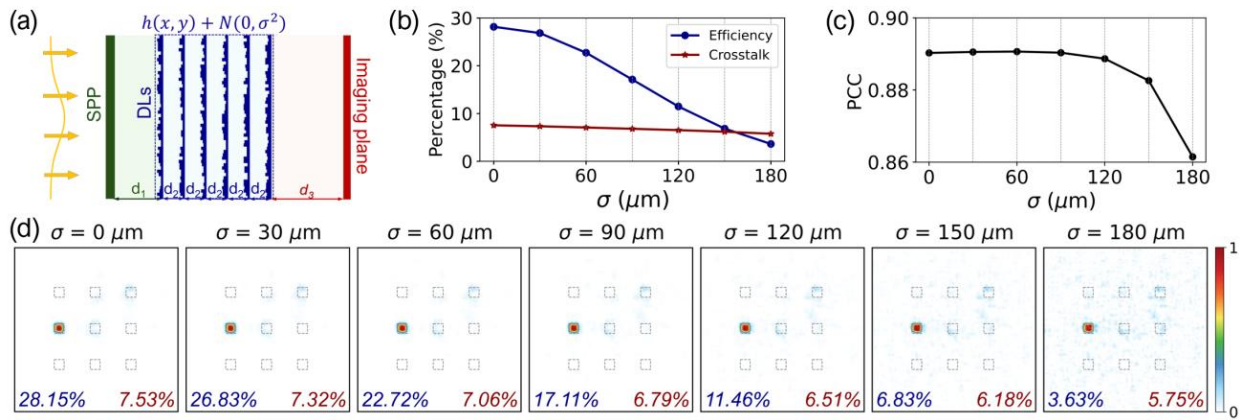

**Fig. 9.** (a) Schematic of sensitivity analysis for variation in pixel heights. (b) Conversion efficiency and crosstalk of $SPP_4$ under different pixel height noise levels. (c) PCC values versus pixel height noise. (d) Imaging plane intensity distributions of $SPP_4$ under varied standard deviation of zero-mean Gaussian random height noise.

Although the DONN is designed to map individual OAM modes to a predefined spatial location, it should also function for the coherent superposition of OAM modes, where the electric fields of each mode are summed. To evaluate this capability, seven different mode superpositions are exemplarily shown: 1&5, 1&6, 2&4, 3&6, 2&3&8, 3&6&8, and 1&2&6&8. Their corresponding phase distributions are shown in Fig. 10(a). The associated SPPs were fabricated via 3D printing using HIPS filament, and photographs of the printed SPPs are shown in Fig. 10(b). The simulated intensity profiles at the imaging plane, shown in Fig. 10(c), demonstrate that all tested OAM beam combinations achieve conversion efficiencies exceeding 21%, confirming the DONN's effectiveness in spatial mode mapping. The measured intensity profiles in Fig. 10(d) closely match the simulations, validating the performance of the fabricated DONN.

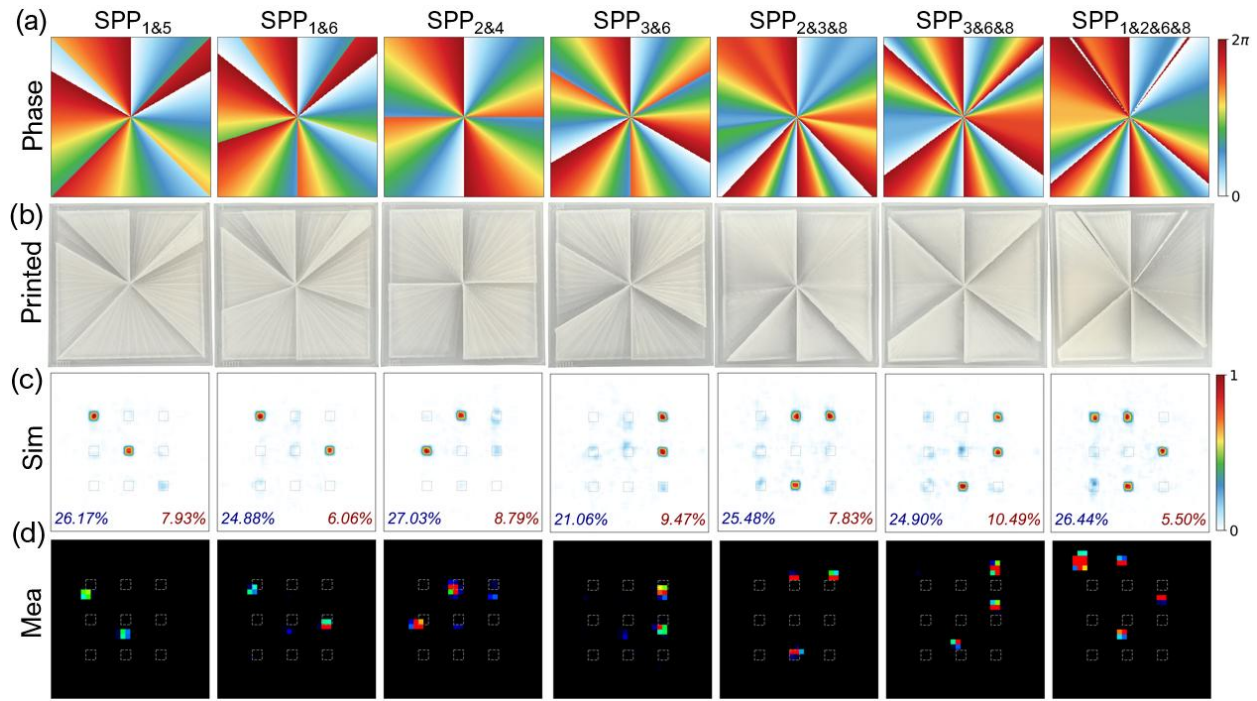

**Fig. 10.** (a) Phase distributions of SPPs for the superposition of OAM beams. (b) Corresponding 3D printed SPPs. (c) Simulated intensity distribution at the imaging plane. (d) Measured intensity distribution at the imaging plane.

It is worth noting that achieving accurate mapping for all possible OAM mode combinations is fundamentally constrained by the design and fabrication limits of the DONN. First, the relatively large pixel size reduces the network's spatial sampling resolution, thereby limiting the highest spatial frequency components that can be encoded in the phase profile. Second, the finite pixel count of each DL imposes a limit on the spatial bandwidth product (SBP) of DONN, which in turn constrains the number of independent spatial channels that can be resolved without crosstalk. Additionally, the six-layer configuration restricts the network's transformation complexity and limits its ability to fully discriminate among OAM superpositions. Enhancing performance requires either increasing the SBP by reducing pixel size and enlarging the aperture, potentially via higher-resolution fabrication or adding more DLs to expand the transformation space. Furthermore, implementing multiple DLs necessitates low-loss or ideally low-absorption 3D printing materials transparent in the target THz spectral range. Otherwise, absorption losses will degrade the overall network performance.

## 5. Conclusion

In this work, we have demonstrated a compact and efficient six-layer DONN for the spatial mapping and recognition of discrete and superposed OAM modes in the THz regime. Utilizing inverse design and a low-cost 3D printing process with HIPS, the DONN successfully distinguishes nine fundamental OAM states with TCs ranging from 1 to 9, as well as multiple superpositions of these modes. Experimental measurements at 0.3 THz confirm the good fidelity in mode identification, showing strong agreement with the simulated intensity profiles. The network achieves good conversion efficiency with low crosstalk, demonstrating robust mode separation and mapping capabilities. Furthermore, we have analyzed the robustness of the DONN against imaging distance variations and fabrication imperfections, showing stable performance within practical tolerances. Constraints imposed by pixel size, aperture dimension, and a limited number of diffractive layers suggest opportunities for future improvements through higher-resolution fabrication, larger apertures, and multi-layer architectures with the selection of low-loss materials transparent at THz frequencies. Overall, this study establishes a scalable, cost-effective platform for THz OAM detection and spatial multiplexing and demultiplexing, paving the way for advanced applications in high-capacity wireless communications, imaging, and sensing that exploit the unique properties of OAM in the terahertz domain.

**Acknowledgements**

The authors acknowledge funding support from the National Science Foundation (# 2420706, #2234413).

**Conflict of Interest**

The authors declare no conflict of interest.

**Data Availability Statement**

All the data and methods needed to evaluate the conclusions in this work are presented in the main text and the Supplementary Information. Any other relevant data is available from the authors upon reasonable request.

**Code Availability**

The code used in this work is based on publicly available standard libraries, primarily implemented using TensorFlow.